\documentclass[11pt,letterpaper]{article}

\usepackage[T1]{fontenc}
\usepackage[utf8]{inputenc}

\usepackage{newtxtext}
\usepackage{newtxmath}
\usepackage[scaled=0.92]{helvet}

\usepackage[margin=1in]{geometry}
\usepackage{graphicx}
\usepackage{microtype}
\usepackage[font=small,labelfont={sf,bf},labelsep=colon]{caption}
\usepackage{enumitem}
\setlist[itemize]{itemsep=0pt,topsep=4pt}

\usepackage{titlesec}
\titleformat*{\section}{\sffamily\bfseries\large}
\titleformat*{\subsection}{\sffamily\bfseries\normalsize}

\usepackage[colorlinks=true,
            linkcolor=blue,
            citecolor=blue,
            urlcolor=blue]{hyperref}

\hypersetup{
  pdfauthor = {Niklas Elmqvist},
  pdftitle = {Embodied Human-Data Interaction},
  pdfsubject = {embodied interaction},
  pdfkeywords = {visualization, visual analytics, sensemaking, embodiment},
}

\usepackage[numbers,sort&compress]{natbib}
\usepackage{doi}

\begin{document}

\begin{center}
  {\sffamily\bfseries\LARGE Embodied Human-Data Interaction\par}
  \vspace{1.5em}
  {\sffamily
    {\large\bfseries Niklas Elmqvist}\\[0.3em]
    Department of Computer Science, Aarhus University\\
    Aarhus, Denmark\\
    \href{mailto:elm@cs.au.dk}{elm@cs.au.dk}\par}
  \vspace{1.5em}
\end{center}

\suppressfloats[t]

\begingroup
\renewcommand{\thefootnote}{}
\footnotetext{Originally published in the \emph{Embodied Interaction: Theory and Practice in HCI} workshop at the ACM SIGCHI Conference on Human Factors in Computing Systems (CHI 2011), May 7--12, 2011, Vancouver, BC, Canada.
  Workshop DOI: \href{https://doi.org/10.1145/1979742.1979592}{10.1145/1979742.1979592}.
  At the time of original publication, the author was with Purdue University, West Lafayette, IN, USA.
  This arXiv version (November 2026) updates the author affiliation and adds article DOIs; the content is otherwise unchanged.}
\endgroup

\noindent{\sffamily\bfseries Abstract}\par
\noindent
We propose the concept of \textit{human-data interaction} (HDI) as the human manipulation, analysis, and sensemaking of large, unstructured,
and complex datasets.
We then proceed to suggest an \textit{embodied} approach to HDI to build physical reasoning environments for these difficult and polymorphic tasks. 
We illustrate our approach with two examples: tangible lenses for visual exploration, and an educational computer game with tangible bricks.


\section{Introduction}

\begin{figure}[t]
  \centering
  \includegraphics[width=0.75\textwidth]{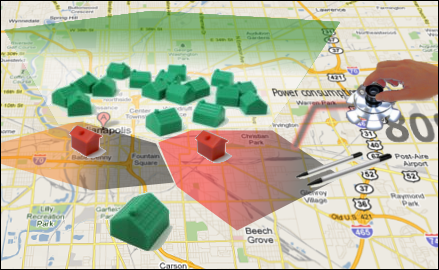}
  \caption{Mockup image of our educational urban planning game.}
  \label{fig:environopoly}
\end{figure}

Information is without a doubt the single most valuable commodity of the new millenium.
It drives the world's financial markets, it imparts knowledge across all Earth's sentient inhabitants, and it governs the ebb and tide of human conflict around the globe.
Just like any other commodity, information does not spring into existence of its own accord, but must be \textit{extracted} from raw data, similar to how petroleum products must be refined from crude oil to be useful.
However, where oil refinement is a tried and tested process, extracting information from raw data is still a largely unmapped and idiosyncratic process that is far from being standardized, widely known, and automated.
For this reason, \textit{data}---and the extraction of information from data---can be seen as the next frontier for the field of computing and, indeed, society as a whole.

While many separate terms exist for these analytical activities, we propose to call them by a single name: \textit{human-data interaction} (HDI), defined as the human manipulation, analysis, and sensemaking~\cite{Russell1993} of data.
Realistic human-data interaction tasks typically exhibit one or several of the following characteristics:

\begin{itemize}
\item\textbf{Large scale.} Datasets are on the order of thousands, millions, or even billions of items.
\item\textbf{Unstructured.} The data is often heterogeneous and lacks coherent structure (e.g., binary).
\item\textbf{Multiple sources.} The data must be marshalled from several different sources, each with their own data format and with no explicit relations.
\item\textbf{Complex relations.} Analysis requires synthesis and correlation, not just queries or computation.
\end{itemize}

Clearly, all of these characteristics contribute to making human-data interaction an extremely challenging task that typically requires analysts to operate at their highest level of performance.
In this position paper, we propose to enable analysts to reach this level of performance by applying embodied interaction~\cite{Dourish2001} to HDI, giving rise to what we call \textit{embodied human-data interaction}.
The motivation for such an approach comes from ethnographic studies~\cite{Isenberg2008, Robinson2008} of both casual~\cite{Pousman2007} and expert users alike working in both laboratory settings and ``in the wild'': the information analysis process is a highly iterative and polymorphic workflow characterized by a plethora of external visual, tangible, and cognitive aids such as physical artifacts~\cite{Ishii1997}, annotations, multiple views~\cite{Grudin2001}, environmental cues~\cite{Robinson2008}, and arrangements in time and space~\cite{Isenberg2008, Robinson2008}.
So far, most HDI tools rely on standard computing hardware such as monitor, mouse, and keyboard, and very few tools exist that capitalize on these additional aspects of the analysis process.

\section{Related Work}

\textit{Sensemaking} is the process of finding a mental representation of a data collection in response to a particular problem or task~\cite{Russell1993}.
Sensemaking is characterized by the synthesis of \textit{new} information---\textit{insight}---from the data.
Two primary approaches exist for sensemaking: one driven by computation (such as data mining, cluster analysis, and inferential statistics), and the other driven by visual representations (visualization).
The emerging field of \textit{visual analytics}~\cite{Thomas2005} combines both approaches.

However, as stated above, most existing visual analytics tools remain fettered to the WIMP paradigm and make little use of novel computing hardware beyond a mouse, keyboard, and monitor.
Exceptions include a few recent tools that are beginning to utilize tabletop displays (e.g.,~\cite{Isenberg2007, Kim2010}), tangible mixing-board interfaces~\cite{Crider2007}, haptics and force
feedback~\cite{Paneels2010}, sonification~\cite{Zhao2008}, and multisensory perception~\cite{Loftin2003}.
Bowman et al.~\cite{Bowman2003} propose the notion of \textit{information-rich virtual environments} (IRVEs) as the intersection of information visualization and virtual reality, but limit its scope to virtual 3D worlds and not the real world.

\section{Embodiments for Sensemaking}

The key cognitive mechanism of sensemaking is \textit{external cognition}~\cite{Norman1994}, where physical representations in the real world are used to augment the internal cognitive processes of the human.
Graphical representations are particularly effective for this purpose~\cite{Scaife1996}, which serves as the basic motivation for the fields of visualization and visual analytics.
However, additional modalities of external cognition exist beyond visual representations:

\begin{itemize}
\item\textbf{Spatial arrangements.} Physical space is an important component of sensemaking for external memory~\cite{Andrews2010}, environmental cues~\cite{Robinson2008}, and partitioning~\cite{Grudin2001}.
\item\textbf{Physical artifacts.} Tangible objects provide embodiments of data and computation in the world~\cite{Dourish2001, Ishii1997}.
\item\textbf{Natural interaction.} Engaging the entire human body will enable natural and intuitive interaction~\cite{Jacob2008}.
\end{itemize}

More specifically, an \textit{embodied} approach to human-data interaction will enable us to harness one or several of the above modalities when building sensemaking tools.
This will support a fluid sensemaking process when viewing, manipulating, and analyzing complex datasets.

\section{Application Examples}

Our EHDI approach is still a largely theoretical exercise, but we have built some simple EHDI tools based on these principles: a tangible lens framework and an educational urban planning simulation using
tangible bricks.

\begin{figure}[t]
  \centering
  \includegraphics[width=\textwidth]{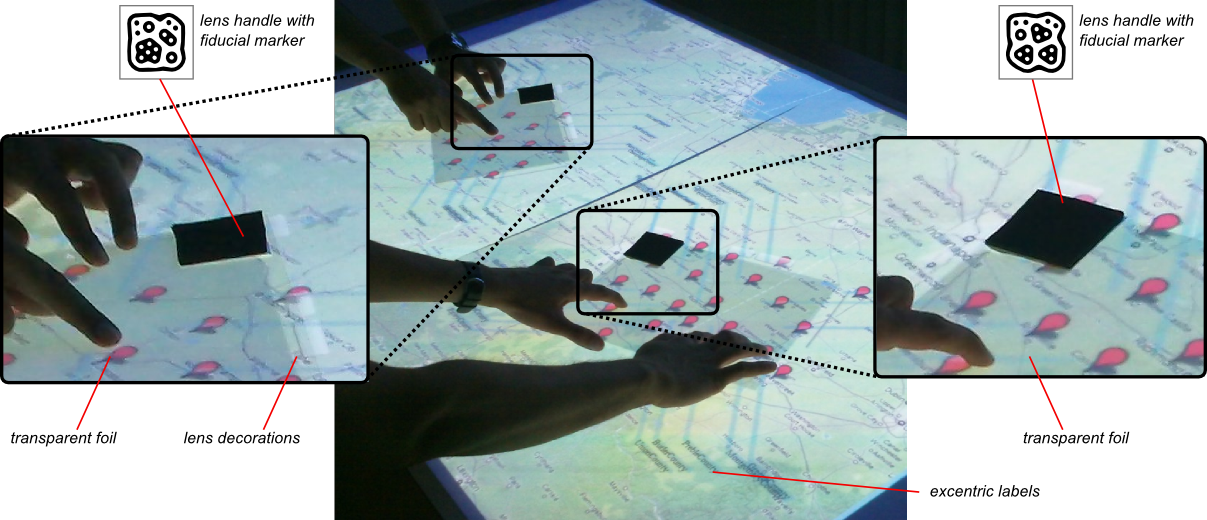}
  \caption{Tangible lens framework: lens handles with fiducial (amoeba)
    markers on transparent foils, with detail insets, on the Google
    Maps application.}
  \label{fig:tanglens}
\end{figure}

\subsection{Tangible and Composable Lenses}

Our framework for tangible and composable lenses combines tabletop displays with tangible lenses as physical embodiments of virtual lenses in graphical space (Figure~\ref{fig:tanglens}).
The lenses are simply transparent sheets of varying shape and size equipped with fiducial markers that allow the tabletop to track their position and rotation.
This data is passed on to the underlying application.
In particular, the physical affordances of these tangible lenses directly suggest that they can be fully or partially overlapped in physical space, causing lens \textit{composition} in virtual space.
We have conducted a formative evaluation studying the semantics of overlapping filtering data.

Using this basic lens framework, we have studied several domain applications of its use.
Figure~\ref{fig:tanglens} shows a map application where users are exploring different data layers using Google Maps as the cartographic dataset.
Each lens reveals measurements within a particular value range for a specific data layer; overlapping lenses yields the union or intersection of results inside the composition.
We have also built an image manipulation application, a 2D scatterplot, and a parallel coordinate plot.

\subsection{Urban Planning Simulation}

\textit{Environpoly} is an educational urban planning simulation that combines tangible bricks with a tabletop display (Figure~\ref{fig:environopoly}).
Essentially a combination of the SimCity computer game and the popular Monopoly board game, the simulation is intended for deployment in a children's museum and designed to teach visitors about sustainable cities and the environmental impact of urban growth.
While it is still in early development, our ambition is to combine a square tabletop display with Monopoly-like tangible bricks for placing residential, commercial and industrial zones; tangible pens to draw roads, power, and water; and physical widgets to control the simulation.

The key aspect of the Environpoly exhibit is the use of embodied interaction to expose the large amount of heterogeneous data that the simulation maintains.
This is particularly important given the intended setting---a museum floor---and the intended user group---children.
For this purpose, we plan on arranging LCD displays around the exhibit that show visual representations of various metrics over time, such as energy usage, pollution, average commute, waste, sewage, and average income.
Furthermore, we also hope to incorporate additional multimodal displays into the exhibit, such as ambient traffic noise to indicate the average level of traffic in the city, as well as a steam generator
disguised as a small smoke stack to indicate the amount of pollution.

\section{Conclusions and Future Work}

We have proposed a novel approach to sensemaking based on applying embodied interaction to the analytical sensemaking of data.
We have also presented two examples of ongoing research that attempts to capitalize on this approach.
However, much work remains in order to fully exploit the ideas outlined in this paper.

\bibliographystyle{abbrvnat}
\bibliography{ehdi-arxiv}

\end{document}